\documentclass[twocolumn]{aastex62}
\submitjournal{The Astrophysical Journal}
\accepted{March 15, 2019}

\usepackage{amsmath, amsthm, amssymb, commath, graphicx, bbold, endnotes, graphicx, subfigure, multirow}

\begin{document}
\title{Fundamental Limitations on the Calibration of Redundant 21~cm Cosmology Instruments and Implications for HERA and the SKA}

\author{Ruby Byrne}
\affiliation{Physics Department, University of Washington, Seattle, WA, USA}

\author{Miguel F. Morales}
\affiliation{Physics Department, University of Washington, Seattle, WA, USA}

\author{Bryna Hazelton}
\affiliation{Physics Department, University of Washington, Seattle, WA, USA}
\affiliation{eScience Institute, University of Washington, Seattle, WA, USA}

\author{Wenyang Li}
\affiliation{Department of Physics, Brown University, Providence, RI, USA}

\author{Nichole Barry}
\affiliation{School of Physics, The University of Melbourne, Parkville, VIC, Australia}
\affiliation{ARC Centre of Excellence for All Sky Astrophysics in 3 Dimensions (ASTRO 3D), Australia}

\author{Adam P. Beardsley}
\affiliation{The School of Earth and Space Exploration, Arizona State University, Tempe, AZ, USA}

\author{Ronniy Joseph}
\affiliation{ARC Centre of Excellence for All Sky Astrophysics in 3 Dimensions (ASTRO 3D), Australia}
\affiliation{International Centre for Radio Astronomy Research  (ICRAR), Curtin University, Bentley WA, Australia}

\author{Jonathan Pober}
\affiliation{Department of Physics, Brown University, Providence, RI, USA}

\author{Ian Sullivan}
\affiliation{Astronomy Department, University of Washington, Seattle, WA, USA}

\author{Cathryn Trott}
\affiliation{ARC Centre of Excellence for All Sky Astrophysics in 3 Dimensions (ASTRO 3D), Australia}
\affiliation{International Centre for Radio Astronomy Research  (ICRAR), Curtin University, Bentley WA, Australia}

\begin{abstract}

Precise instrument calibration is critical to the success of 21~cm cosmology experiments. Unmitigated errors in calibration contaminate the Epoch of Reionization (EoR) signal, precluding a detection. \citealt{Barry2016} characterizes one class of inherent errors that emerge from calibrating to an incomplete sky model, however it has been unclear if errors in the sky model affect the calibration of redundant arrays. In this paper, we show that redundant calibration is vulnerable to errors from sky model incompleteness even in the limit of perfect antenna positioning and identical beams. These errors are at a level that can overwhelm the EoR signal and prevent a detection on crucial power spectrum modes. Finally, we suggest error mitigation strategies with implications for the Hydrogen Epoch of Reionization Array (HERA) and the Square Kilometre Array (SKA).

\end{abstract}



\section{Introduction}

The promise of 21~cm cosmology observations to provide crucial constraints on the Epoch of Reionization (EoR) and Dark Energy depends on experimenters' ability to suppress the bright astrophysical foregrounds. The astrophysical foregrounds are 4--5 orders-of-magnitude brighter than the faint cosmological signal, but because they are spectrally smooth, they are in principle separable. For perfectly calibrated instruments this leads to the current paradigm of 21~cm power spectrum (PS) observations where there is a contaminated `foreground wedge' and a clean `measurement window' or `EoR window.' (See \citealt{Furlanetto2006b} and \citealt{Morales2010} for general reviews and \citealt{Datta2010,Morales2012,Vedantham2011,Parsons2012,Trott2012,Hazelton2013,Thyagarajan2013,Dillon2013}; and \citealt{Thyagarajan2015} for discussions of how smooth spectrum foregrounds appear in PS measurements.) 

However, it has been recognized for some time that small chromatic instrumental calibration errors limit the separability of the cosmological signal and bright foreground emission. When applied to data, frequency-dependent errors in calibration introduce frequency structure into the spectrally smooth foregrounds. This couples foreground power into the EoR window and can overwhelm the faint cosmological PS. Precision calibration is therefore critical for enabling 21~cm cosmology, and characterizing and mitigating calibration errors has become a very active area of research \citep{Pen2009,grobler2014,Newburgh2014,Berger2016,grobler2016,Patil2016,vanweeren2016,wijnholds2016,Ewall-Wice2017,Joseph2018,Li2018,orosz_2018}. 

One such mechanism of error was identified in \citealt{Barry2016}. This sky-based calibration error emerges from fitting antenna bandpasses to an incomplete sky model. Missing sources in the sky model introduce errors in the calibration solutions. Because of instruments' frequency-dependent point spread functions (PSFs), these errors are chromatic and couple foreground power into the EoR window.

\citealt{Barry2016} describes this mechanism in the context of traditional sky-based calibration. However, to achieve the extraordinary calibration precision needed for 21~cm cosmology many efforts have been designed around redundant arrays and calibration schemes (e.g.\ \citealt{wieringa_redundant_cal,Parsons2010,Dillon2016,DeBoer2017,dillon_polarized_cal,grobler2018}). The community has often assumed that redundant calibration approaches are immune to the effect described in \citealt{Barry2016} and similar systematics.

This paper is part of an ongoing exploration of the limits of redundant calibration. \citealt{Joseph2018} and \citealt{orosz_2018} theoretically examine the effects of antenna non-redundancy due to position offsets and beam irregularities. \citealt{Li2018} uses the unique array layout of Phase II of the MWA \citep{mwa_phase_II_design} to produce a direct comparison of precision sky-based and redundant calibration solutions.

In this paper we show that the calibration errors associated with an incomplete sky model affect redundant calibration through the absolute calibration step. Even in the limit of a perfectly redundant array with identical antenna beam responses, sky model incompleteness introduces frequency-dependent calibration errors that contaminate the PS measurement.

In \S\ref{section_calibration} we develop our mathematical framework and identify the channels through which sky model errors affect redundant calibration. In \S\ref{section_errors} we repeat the simulations of \citealt{Barry2016} for a hexagonal array. We show that the systematics identified in \citealt{Barry2016} affect the hexagonal array's redundant calibration solutions. In \S\ref{section_layouts} we extend our simulations to multiple redundant and non-redundant arrays to explore the effect of array layout and demonstrate that calibration errors from sky model incompleteness are typically worse for the regular arrays required by redundant calibration. In \S\ref{section_discussion} and \S\ref{section_conclusion} we discuss the impact of the systematics identified in this paper and propose error mitigation strategies with implications for HERA (the Hydrogen Epoch of Reionization Array) and the SKA (Square Kilometer Array).


\section{Calibration Formalism} \label{section_calibration}

Calibration is an integral aspect of radio interferometry when the instrument response varies across individual array antennas, frequency channels, or measurement times. While there are innumerable calibration strategies and algorithms, calibration methods can be broadly categorized as sky-based, redundant, hybrid, or external. Sky-based calibration uses a sky model as a prior and is well-suited to imaging arrays with good UV coverage. Redundant calibration works for highly regular arrays and calibrates by matching visibilities from redundant baselines \citep{wieringa_redundant_cal, Liu2010}. Hybrid calibration combines elements of both sky-based and redundant calibration and is a nascent area of study \citep{Li2018,sievers2017}. External calibration uses a separate calibration source such as a drone \citep{Jacobs2017}, satellites \citep{Neben2015,Neben2016a}, an injected noise source \citep{Newburgh2014}, or a pulsar to measure the antenna response \citep{Pen2009}.

Understanding the mathematical framework for each calibration technique is critical for identifying sources of calibration error. Here we discuss sky-based and redundant calibration and identify the channels through which sky model incompleteness introduces errors in the calibration solutions.

\subsection{The Measurement Equation}

The basis of most interferometric calibration is the measurement equation, which relates the measured sky visibility $v_{jk}(f)$ from antennas $j$ and $k$ at frequency $f$ to the theoretical `true' sky visibility for that baseline and frequency, $u_{jk}(f)$:
\begin{equation}
    v_{jk}(f) = G_{jk}(f) u_{jk}(f) + n_{jk}(f)
\end{equation}
\citep{Hamaker1996}. Here $G_{jk}(f)$ is the instrument gain and $n_{jk}(f)$ is the noise. 

Each term in the measurement equation is implicitly per-time. The antenna indices $j$ and $k$ index each polarization of each antenna. In this paper we consider just one polarization mode to simplify the analysis with no loss of generality.

Different calibration approaches correspond to different parameterizations of $G_{jk}(f)$, and finding the optimal parameterization is a central challenge of precision calibration. Traditional calibration assumes per-antenna and per-frequency gains, such that $G_{jk}(f) = g_j(f) g_k^*(f)$ where $g_j(f)$ is the gain for antenna $j$ at frequency $f$.

\subsection{Sky-Based Calibration}

Sky-based calibration uses a sky model as a prior. The sky model is simulated through a model of the instrument to derive model visibilities, $m_{jk}(f)$. A calibration solution is calculated by approximating the true sky visibilities with the model visibilities, $u_{jk}(f) \approx m_{jk}(f)$. The measurement equation then becomes
\begin{equation}
    v_{jk}(f) \approx g_j(f) g_k^*(f) m_{jk}(f) + n_{jk}(f).
\end{equation}

Assuming the noise is Gaussian, mean-zero, and uncorrelated with variance $\sigma_{jk}^2(f)$, a maximum likelihood estimate of the per-antenna gains maximizes
\begin{equation}
\begin{split}
    \mathcal{L}&(\{g_j(f)\} | \{v_{jk}(f)\}, \{m_{jk}(f)\}) \\
    & \propto \prod_f \prod_{jk} e^{-\frac{1}{2}\left(\frac{|g_j(f) g_k^*(f) m_{jk}(f)-v_{jk}(f)|}{\sigma_{jk}(f)}\right)^2}.
\end{split}
\end{equation}
By taking the logarithm of both sides, we find that maximizing $\mathcal{L}$ is equivalent to minimizing the per-frequency $\chi$-squared by varying the per-antenna, per-frequency gains ${g_j(f)}$:
\begin{equation}
    \chi^2_{\text{sky}}(f) = \sum_{jk} \frac{| v_{jk}(f) - g_j(f) g_k^*(f) m_{jk}(f) |^2}{\sigma_{jk}^2(f)}.
\label{sky_cal_x2}
\end{equation}
These calculated gains are denoted $\hat{g}_j(f)$, where the `hat' symbol indicates the maximum-likelihood estimate.

\subsection{Redundant Calibration}

Redundant calibration works for highly regular arrays with many redundant baselines. It calibrates by imposing a prior that true sky visibilities from redundant baselines are equal \citep{wieringa_redundant_cal,Liu2010}. Instead of approximating the true sky visibilities with a model, redundant calibration solves for the true sky visibilities for each redundant baseline set alongside the gains. Highly regular arrays have many more measured visibilities than unique baselines, so the system is overdetermined even when treating the sky visibilities as free parameters.

The measurement equation for redundant calibration replaces the sky visibilities, $u_{jk}(f)$, with visibility terms that are constrained to be equal across redundant baselines, $u_{\alpha}(f)$ where $\alpha$ indexes the redundant baseline sets:
\begin{equation}
    v_{jk}(f) \approx g_j(f) g_k^*(f) u_{\alpha}(f) + n_{jk}(f).
\end{equation}
As in sky-based calibration, we assume Gaussian, uncorrelated noise and construct a maximum-likelihood estimate for the gains and sky visibilities. Here the likelihood function is
\begin{equation}
\begin{split}
    \mathcal{L}&(\{g_j(f)\}, \{u_{\alpha}(f)\}| \{v_{jk}(f)\}) \\ 
    &\propto \prod_f \prod_{\alpha} \prod_{\{j, k\}_{\alpha}} e^{-\frac{1}{2}\left(\frac{|g_j(f) g_k^*(f) u_{\alpha}(f)-v_{jk}(f)|}{\sigma_{jk}(f)}\right)^2},
\end{split}
\end{equation}
where $\{j, k\}_{\alpha}$ are the sets of antennas that belong to each redundant baseline type $\alpha$. Maximizing this function is equivalent to minimizing
\begin{equation}
\begin{split}
    \chi^2_{\text{red}}(f) &=  \sum_{\alpha} \chi_{\alpha,\text{red}}^2(f) \\
    &= \sum_{\alpha} \sum_{\{j, k\}_{\alpha}} \frac{| v_{jk}(f) - g_j(f) g_k^*(f) u_{\alpha}(f) |^2}{\sigma_{jk}^2(f)}
\end{split}
\label{red_x2}
\end{equation}
by varying $g_j(f)$ and $u_{\alpha}(f)$ for each frequency $f$.

However, minimizing Equation \ref{red_x2} yields degenerate solutions \citep{Liu2010}. The degeneracies can be parameterized as four terms per frequency: overall amplitude $A(f)$, overall phase $\Delta(f)$, and two phase gradient components $\Delta_x(f)$ and $\Delta_y(f)$. Transformations of these parameters leave $\chi^2_{\text{red}}(f)$ unchanged (here we have omitted explicit frequency dependence):
\begin{itemize}
    \item Overall amplitude $A$: The transformation $g_j \rightarrow A g_j$ does not change the form of $\chi^2_{\text{red}}$ if it is accompanied by the transformation $u_{\alpha} \rightarrow A^{-2} u_{\alpha}$. Errors in the overall amplitude change the sky brightness, making the sky appear artificially bright or dim.
    \item Overall phase $\Delta$: The transformation $g_j = |g_j|e^{i\phi_j} \rightarrow |g_j|e^{i(\phi_j + \Delta)}$ corresponds to $g_j g_k^* = |g_j| |g_k| e^{i(\phi_j - \phi_k)} \rightarrow |g_j| |g_k| e^{i(\phi_j + \Delta - \phi_k - \Delta)} = g_j g_k^*$, so $\chi^2_{\text{red}}$ is unchanged under this transformation. Note that is is also true for $\chi^2_{\text{sky}}$ from Equation \ref{sky_cal_x2}; this degeneracy exists in sky calibration as well as redundant calibration.
    \item Phase gradient $\Delta_x$ and $\Delta_y$: Assuming a co-planar array, the transformation $g_j = |g_j|e^{i\phi_j} \rightarrow |g_j|e^{i(\phi_j + \Delta_x x_j + \Delta_y y_j)}$ does not change the form of $\chi^2_{\text{red}}$ if it is accompanied by the transformation $u_{\alpha} = |u_{\alpha}| e^{i \phi_{\alpha}} \rightarrow |u_{\alpha}| e^{i (\phi_{\alpha} - \Delta_x x_{\alpha} - \Delta_y y_{\alpha})}$. Here $(x_j, y_j)$ are the $x$- and $y$-coordinates of the position of antenna $j$, and $(x_{\alpha}, y_{\alpha})$ are the $x$- and $y$-separations of antennas that form baselines in redundant baseline set $\alpha$. Errors in the phase gradient parameters shift the sky image such that sources appear offset from their true positions.
\end{itemize}

Additional degeneracies arise in special cases. Arrays with separate redundant sub-arrays can have more than four degenerate parameters per frequency; for example, each of the hexagonal sub-arrays in the MWA Phase II has an independent overall phase degeneracy \citep{Li2018}. Furthermore, while this paper assumes the simple case of single-polarization calibration, fully polarized redundant calibration has degeneracies associated with the coupling between polarizations \citep{dillon_polarized_cal}.

Redundant calibration must be separated into two distinct parts because of the degeneracies of solutions that minimize Equation \ref{red_x2}. `Relative calibration' solves for the antenna gains up to the degenerate parameters $A(f)$, $\Delta(f)$, $\Delta_x(f)$, and $\Delta_y(f)$, and `absolute calibration' constrains those degeneracies. We can parameterize the redundant calibration solutions to reflect the inherent separation of redundant calibration into relative and absolute calibration steps. We define a set of parameters $h_j(f)$ to be the gains constrained to have an average amplitude of 1, average phase of 0, and phase gradient of 0. Now 
\begin{equation}
    g_j(f) = A(f) e^{i[\Delta(f) + \Delta_x(f) x_j + \Delta_y(f) y_j]} h_j(f)
\label{gains_decomp}
\end{equation}
where $(x_j, y_j)$ is the position of antenna $j$. 

A non-degenerate formulation of Equation \ref{red_x2} is
\begin{equation}
\begin{split}
    \chi^2_{\text{red}}(f) &=  \sum_{\alpha} \chi_{\alpha,\text{red}}^2(f) \\
    &= \sum_{\alpha} \sum_{\{j, k\}_{\alpha}} \frac{| v_{jk}(f) - h_j(f) h_k^*(f) w_{\alpha}(f) |^2}{\sigma_{jk}^2(f)},
\end{split}
\label{red_x2_nondegenerate}
\end{equation}
where $w_{\alpha}(f) = A^2(f) e^{i[\Delta_x(f) x_{\alpha} + \Delta_y(f) y_{\alpha}]} u_{\alpha}(f)$. Here $x_{\alpha}$ and $y_{\alpha}$ are the $x$- and $y$-coordinates of baselines in redundant baseline set $\alpha$. Relative calibration minimizes this expression by varying $h_j(f)$ and $w_{\alpha}(f)$.

Absolute calibration solves for $A(f)$, $\Delta(f)$, $\Delta_x(f)$, and $\Delta_y(f)$. These parameters cannot be constrained from baseline redundancy and are generally calculated from a sky model. Here we describe a typical implementation based on fitting the absolute calibration parameters to sky-based calibration solutions. Other absolute calibration methods avoid explicit sky-based calibration by fitting the absolute calibration parameters directly from the model visibilities. HERA Memo \#063\footnote{\texttt{https://reionization.org/science/memos/}} compares two absolute calibration techniques and shows that they yield consistent results.

Minimizing Equation \ref{sky_cal_x2} with model visibilities $m_{jk}(f)$ gives a set of maximum-likelihood estimated sky-based gains $\hat{g_j}^{\text{sky}}(f)$. The overall amplitude can be fit by averaging across the sky-based gain amplitudes:
\begin{equation}
    \hat{A}(f) = \frac{1}{N} \sum_{j=0}^N |\hat{g_j}^{\text{sky}}(f)|.
\label{A_calc}
\end{equation}
The phase gradient parameters can be fit by minimizing the expression 
\begin{equation}
\begin{split}
    \chi_{\phi}^2(f) = \sum_{j=0}^N &\left(\operatorname{Arg}[\hat{g_j}^{\text{sky}}(f)]\right. \\ 
    &\left.-\Delta(f) - \Delta_x(f) x_j - \Delta_y(f) y_j\right)^2,
\end{split}
\label{grad_calc}
\end{equation}
by varying $\Delta$, $\Delta_x$ and $\Delta_y$. Here $\operatorname{Arg}[\hat{g_j}^{\text{sky}}(f)]$ is the complex phase of $\hat{g_j}^{\text{sky}}(f)$. We assume that $\operatorname{Arg}[\hat{g_j}^{\text{sky}}(f)] \ll 2\pi$ and therefore do not have to account for the branch cut in the complex plane. 

The overall phase $\Delta(f)$ is degenerate in Equation \ref{sky_cal_x2}, the $\chi$-squared for sky-based calibration, so it must be calculated in another way. One typical way to set the overall phase for either sky-based or redundant calibration is to use a reference antenna. The overall phase would then be set by requiring that $\operatorname{Arg}[\hat{g}_{\text{ref}}(f)]=0$ where $\hat{g}_{\text{ref}}(f)$ is the gain of the reference antenna. 

Redundant calibration requires combining the relative and absolute calibration steps according to Equation \ref{gains_decomp} to get the true calibration solutions $\{\hat{g_j}(f)\}$.

\subsection{Comparison of Sky-Based and Redundant Calibration}

As shown above, redundant calibration consists of relative and absolute calibration steps. This is motivated by the degeneracies in Equation \ref{red_x2}, but we can also apply the same parameterization to solutions from sky-based calibration. We can decompose gains from sky-based calibration into relative and absolute calibration components:
\begin{equation*}
    g_j(f) = \overbrace{A(f) e^{i[\Delta(f) + \Delta_x(f) x_j + \Delta_y(f) y_j]}}^{\text{Abs. Cal.}} \overbrace{h_j(f)}^{\text{Rel. Cal.}}.
\end{equation*}
The absolute calibration parameters $A(f)$, $\Delta(f)$, $\Delta_x(f)$, and $\Delta_y(f)$ describe the bulk array response across all antennas. The relative calibration parameters $h_j(f)$ fit the calibration degrees of freedom that describe differences between antennas. 

This decomposition allows us to directly compare calibration parameters between sky-based and redundant calibration methods. Calibration error mechanisms can be classified as affecting relative calibration, absolute calibration, or both. The errors depend on the specific calibration methods used and the features of the instrument.

Relative calibration is not necessary for instruments with highly uniform antenna responses. In this regime it is advantageous to require uniformity across all antennas' calibration solutions, preventing overfitting of antenna-to-antenna structure. Averaging per-antenna gains is equivalent to setting $h_j(f)=1$ for all antennas and frequencies (as well as setting $\Delta_x=\Delta_y=0$). \citealt{Barry2016} demonstrates that this averaging mitigates calibration errors from sky model incompleteness (see the green `maximally averaged' line in Figure 8 of \citealt{Barry2016}, which includes both antenna- and time-averaging).

Redundant calibration does not require a sky model for relative calibration, and it excels at fitting antenna-to-antenna variations. At the same time, redundant calibration inherently assumes uniform beam responses across antennas. The fundamental assumption of redundant calibration --- that redundant baselines measure the same sky visibility --- breaks down if the antennas have different beam responses. Therefore redundant calibration is best suited to arrays in which antenna-to-antenna variations occur in the analog signal path after the receiving element. These variations are consistent with redundant calibration's assumptions and can be captured by relative calibration.

Any calibration steps that rely on a sky model are susceptible to errors from an inaccurate or incomplete sky model. Sky-based calibration uses a sky model in both relative and absolute calibration, while redundant calibration uses a sky model in absolute calibration only. Developing better sky models is an active area of research \citep{Carroll2016, Hurley-Walker2017}, but no realistic sky model can achieve perfect accuracy and completeness. As shown in \citealt{Barry2016}, missing sources in the sky model introduce errors in sky-based calibration solutions. This error mechanism affects absolute calibration and therefore impacts redundant calibration.

In this paper we investigate errors introduced in absolute calibration due to sky model incompleteness. These errors are independent of the array redundancy requirement and are present even in the limit of perfect redundancy. For the purposes of this paper we assume perfect relative calibration, i.e.\ no errors in $\hat{h_j}(f)$, in order to focus on the errors in absolute calibration. For discussions of relative calibration errors in redundant calibration, see \citealt{orosz_2018, Li2018}; and \citealt{Joseph2018}.


\section{Errors in Redundant Calibration Due to an Incomplete Sky Model} \label{section_errors}

Redundant calibration is susceptible to errors from an incomplete sky model that enter through the absolute calibration step. In this section, we show that these errors are frequency-dependent and that they contaminate the EoR PS.

\begin{figure}
\centering
\includegraphics[width=\columnwidth]{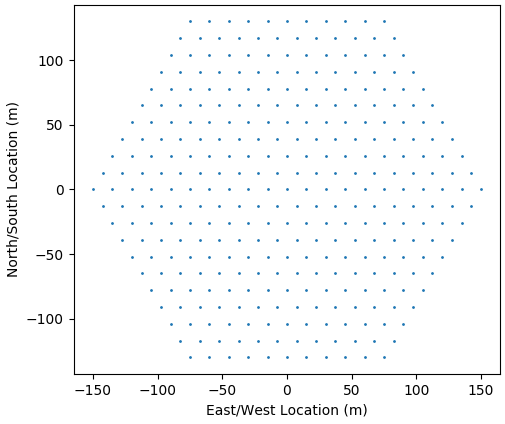}
\caption{Representation of the hexagonal array layout used in simulation. The array has 331 antennas and minimum antenna spacings of 15 m.}
\label{fig: hex antenna layout large}
\end{figure}

We simulate these errors for a redundant array of 331 antennas, arranged in a regular hexagonal layout with minimum antenna spacings of 15 m (see Figure \ref{fig: hex antenna layout large}). The simulations use the MWA antenna zenith pointing beam model in the 167-198 MHz frequency band. One beam model is used for all frequencies across this band to eliminate errors from frequency-dependent beam modulation. For simplicity, we consider only one polarization; all results in this paper use simulated data from East-West dipole antennas. Visibilities are created for a 2-minute observation with the Fast Holographic Deconvolution (FHD) software pipeline\footnote{\texttt{https://github.com/EoRImaging/FHD}} \citep{Sullivan2012} and are based on the GLEAM catalog \citep{Hurley-Walker2017} at the `EoR-0' field (centered on Right Ascension 0.00 h and Declination $-27^{\circ}$) for a total of 51,821 simulated sources with a minimum flux density of 10 mJy. We then create a calibration catalog from the 4,000 brightest sources in apparent flux density (minimum flux density 89 mJy), as was done in \citealt{Barry2016}. By calibrating on only a subset of the simulated catalog, we represent the fact that calibration catalogs are realistically incomplete. The missing sources in the calibration catalog introduce errors in the calibration solutions.

To calibrate, we first implement sky-based calibration with FHD. This minimizes the $\chi^2_{\text{sky}}$ from Equation \ref{sky_cal_x2}, where the `measured visibilities' $v_{jk}$ are simulated from 51,821 sources and the `modeled visibilites' $m_{jk}$ are simulated from 4,000. This gives per-antenna, per-frequency gain solutions $\hat{g_j}^{\text{sky}}(f)$. We then calculate the absolute calibration solutions from those gains. 

\begin{figure}
\centering
\includegraphics[width=\columnwidth]{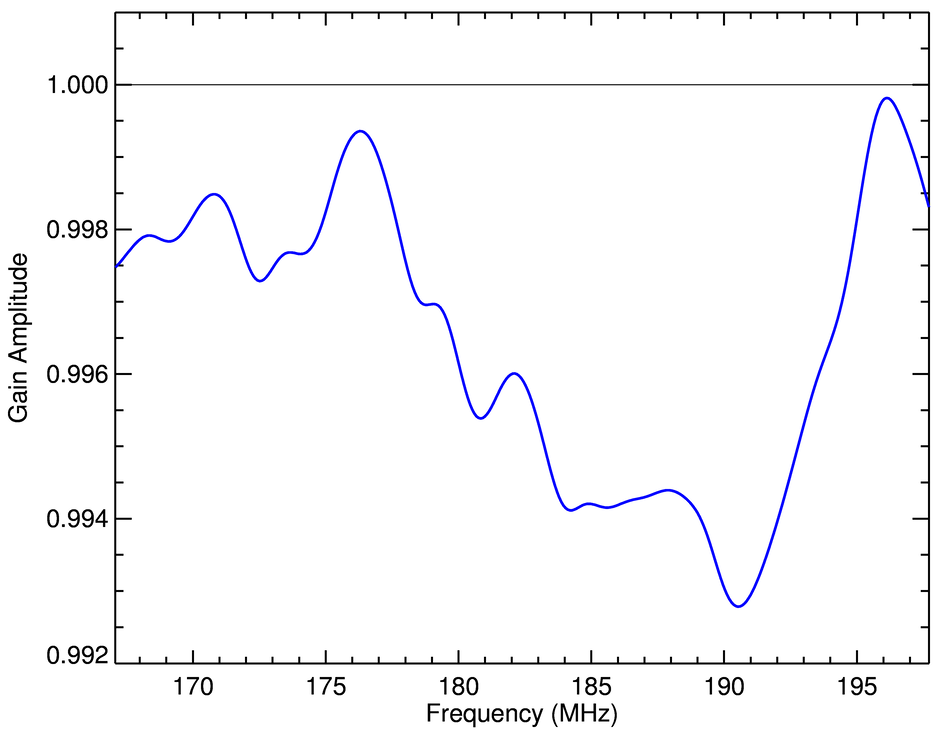}
\caption{Plot of the average amplitude of the per-antenna gains as a function of frequency, $\hat{A}(f)$, for the hexagonal array in Figure \ref{fig: hex antenna layout large}. The visibilities are calibrated to an incomplete catalog that includes only a subset of the simulated sources. Deviations from 1 correspond to calibration errors due to an incomplete sky model. Since the average gain amplitude is a degenerate parameter in the relative calibration step of redundant calibration, these errors persist in redundant calibration even in the limit of perfect redundancy. The specific features of the errors in this parameter depend on the locations and flux densities of the sources missing from the sky model. In this case, those missing sources are faint sources in the `EoR-0' field, as described by the GLEAM catalog \citep{Hurley-Walker2017}.}
\label{fig: hex avg amp errors}
\end{figure}

We use equation \ref{A_calc} to calculate the overall amplitude, $\hat{A}$, plotted in Figure \ref{fig: hex avg amp errors} as a function of frequency. Deviations from $\hat{A}=1$ are calibration errors due to the incompleteness of the sky model. These errors are frequency-dependent. When applied to data, they introduce frequency structure to the intrinsically spectrally smooth foregrounds, coupling their power into the EoR window and obscuring the EoR signal.

\begin{figure}
\centering
\subfigure{
\includegraphics[width=\columnwidth]{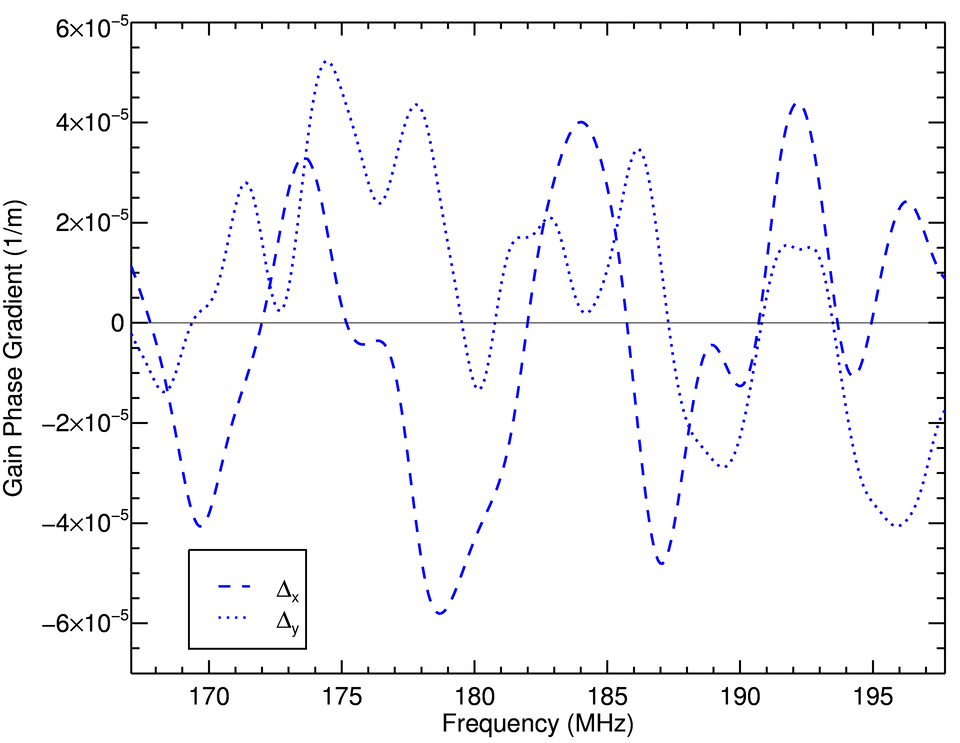}
}
\caption{Plot of the gain complex phase gradient fit parameters as a function of frequency, $\hat{\Delta_x}(f)$ and $\hat{\Delta_y}(f)$, for the simulation described in Figure \ref{fig: hex avg amp errors}. Here $x$ refers to the East-West direction and $y$ refers to the North-South direction. Deviations from 0 correspond to calibration errors due to an incomplete sky model. Like the average gain amplitude, these two phase gradient parameters are degenerate in relative calibration, so these errors persist in redundant calibration even in the limit of perfect redundancy.}
\label{fig: hex array avg phase grad}
\end{figure}

We calculate the phase gradient parameters $\hat{\Delta}_x$ and $\hat{\Delta}_y$, plotted in Figure \ref{fig: hex array avg phase grad}, from Equation \ref{grad_calc}. Here deviations from 0 are calibration errors due to the incompleteness of the sky model. As with the overall amplitude, the phase gradient parameter errors are frequency-dependent and can therefore couple foreground power into the EoR window.

We require that the overall phase $\hat{\Delta}=0$. As $\hat{\Delta}$ is degenerate in both sky and redundant calibration, setting a reference phase is an important aspect of precision calibration across both calibration methods. In simulation the true antenna gain phases are 0, so by setting the overall phase to 0 we simulate perfect calibration of the overall phase.

To demonstrate contamination of the EoR window from the errors in the average gain amplitudes plotted in Figure \ref{fig: hex avg amp errors} and the gain phase gradient fit terms plotted in Figure \ref{fig: hex array avg phase grad}, we produce 2-D PS with the Error Propagated PS with InterLeaved Observed Noise ($\epsilon$ppsilon) software package\footnote{\texttt{https://github.com/EoRImaging/eppsilon}} \citep{Jacobs2016}. The 2-D PS are a function of line-of-sight modes ($k_{\parallel}$, the Fourier modes across frequency) on the vertical axis and modes perpendicular to the line-of-sight ($k_{\bot}$, the Fourier modes across the sky) on the horizontal axis. This 2-D PS space is a useful tool for isolating foreground power and identifying systematics in the analysis pipeline. Here we use this tool to identify power leakage from low to higher $k_{\parallel}$ modes.

\begin{figure*}
\centering
\includegraphics[width=2\columnwidth]{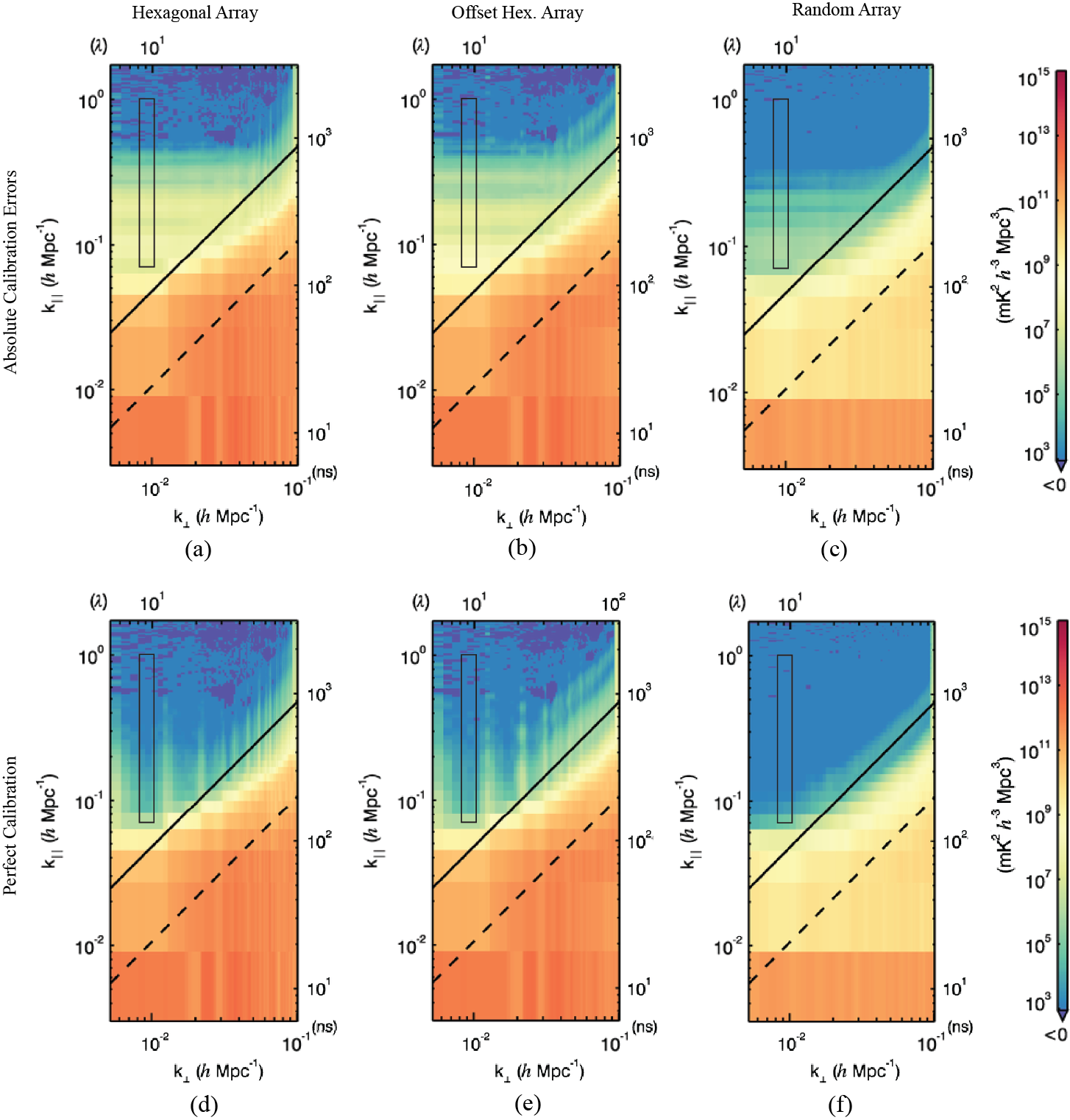}
\caption{2-D `residual' PS of simulated data calibrated to an incomplete sky model. The plots in the top row have been calibrated with errors in the absolute calibration parameters $A$, the overall gain amplitude, and $\Delta_x$ and $\Delta_y$, the gain phase gradient parameters. There are no errors in relative calibration. This corresponds to a regime of perfect array redundancy and no thermal noise, where the only calibration errors emerge from absolute calibration to an incomplete model. The plots in the bottom row have perfect calibration. Notice that errors in the absolute calibration parameters causes power bleed into higher line-of-sight modes. The leftmost column are simulations of the hexagonal array pictured in Figures \ref{fig: hex antenna layout large} and \ref{fig: per antenna array comparison}(a) and discussed in \S \ref{section_errors}. The middle column are simulations from the offset hexagonal array in Figure \ref{fig: per antenna array comparison}(b) and the rightmost column are simulations from the random array pictured in Figure \ref{fig: per antenna array comparison}(c), both discussed in \S \ref{section_layouts}. Line-of-sight modes ($k_{\parallel}$) are plotted on the vertical axis and modes perpendicular to the line-of-sight ($k_{\bot}$) on the horizontal axis. The high power in the lowest line-of-sight ($k_{\parallel}=0$) mode represents the intrinsic foregrounds. The red-orange wedge across the lower right part of the spectrum is the `foreground wedge' and comes from the chromatic instrument response, which mixes the intrinsic foregrounds with higher line-of-sight modes \citep{Datta2010,Morales2012,Vedantham2011,Parsons2012,Trott2012,Hazelton2013,Thyagarajan2013,Dillon2013,Thyagarajan2015}. The solid and dashed diagonal lines are the `horizon' and `primary field of view' lines, respectively. These denote contamination limits based on sources' off-axis positions. The vertical streaks visible in the upper half of Figures (d) and (e) are regions of low UV coverage. The black rectangular outlines in each plot denote the values that contribute to the 1-D plots in Figures \ref{fig: hex 1d ps} and \ref{fig: 1d ps}.}
\label{fig: 2d ps}
\end{figure*}

Figure \ref{fig: 2d ps} gives 2-D PS for three separate simulated arrays. The leftmost column of the figure corresponds to the hexagonal array pictured in Figure \ref{fig: hex antenna layout large}. The middle and rightmost columns correspond to additional array configurations discussed in \S \ref{section_layouts}. The PS of the sky model used in calibration has been subtracted, producing `residual' PS.

In the top left, Figure \ref{fig: 2d ps}(a) gives the 2-D PS of simulated visibilities calibrated with absolute calibration errors. The gains applied to these data are $g_j = \hat{A} e^{\hat{\Delta}_x x_j + \hat{\Delta}_y y_j}$, where the parameters $\hat{A}$, $\hat{\Delta}_x$, and $\hat{\Delta}_y$ take the values calculated from Equations \ref{A_calc} and \ref{grad_calc} and plotted in Figures \ref{fig: hex avg amp errors} and \ref{fig: hex array avg phase grad}. Here relative calibration is perfect: calibration errors come from the absolute calibration parameters only. This represents the best possible redundant calibration achievable with the 4,000-source incomplete sky model in the limit of unrealistically perfect redundancy and the absence of noise. Below, Figure \ref{fig: 2d ps}(d) gives the 2-D PS of the same simulated data without calibration, or with $g_i=1$ for all antennas. In simulation this is equivalent to perfect calibration, or calibration to a complete sky model. 

Figures \ref{fig: 2d ps}(a) and \ref{fig: 2d ps}(d) illustrate inherent PS features. First of all, the high power in the lowest line-of-sight mode ($k_{\parallel}=0$) represents the intrinsic foregrounds. Foreground emission is extremely spectrally smooth; in this simulation, it is perfectly flat. The red-orange wedge across the lower right part of the spectrum is the `foreground wedge' and comes from the chromatic instrument response, which mixes the intrinsic foregrounds with higher line-of-sight modes \citep{Vedantham2011,Morales2012,Trott2012,Parsons2012,Hazelton2013}. The vertical streaks in the upper half of the spectrum are an intrinsic PS feature resulting from regions of low UV coverage (Figure \ref{fig: all arrays avg amp errors}(d) shows the UV coverage of this array). The periodicity of these streaks emerges from the regular layout of antennas in the hexagonal array. Averaging over sufficiently long time intervals can mitigate this effect by leveraging the Earth's rotation to fill in areas of low UV coverage through a process called `UV rotation.'

Comparing the uncalibrated PS simulation in Figure \ref{fig: 2d ps}(d) to the PS simulation with absolute calibration errors in Figure \ref{fig: 2d ps}(a) shows that missing sources in the sky model introduce errors that cause power leakage into high line-of-sight PS modes. Even with perfect relative calibration, the frequency-dependent errors in the absolute calibration parameters plotted in Figures \ref{fig: hex avg amp errors} and \ref{fig: hex array avg phase grad} introduce frequency structure into the spectrally smooth foregrounds. This frequency structure results in foreground power leakage into the PS modes in the EoR-sensitive window, obscuring the faint EoR signal.

The power leakage in the EoR window falls off at large $k_{\parallel}$ values. The maximum contaminated $k_{\parallel}$ mode is proportional to the length of the array's longest baseline. For a widefield array sensitive to emission at the horizon, power leakage occurs at a maximum mode of $b/c$, where $b$ is the length of the longest baseline used in calibration and $c$ is the speed of light. Converting to cosmological units \citep{morales2004}, we expect calibration errors to produce power spectrum contamination for the hexagonal array pictured in Figure \ref{fig: hex antenna layout large} on modes $k_{\parallel} \lesssim 0.58 \ h \text{Mpc}^{-1}$. Limiting calibration to short baselines can restrict foreground leakage to low $k_{\parallel}$ modes, freeing a larger region of the EoR window from contamination \citep{Ewall-Wice2017}. However, it is critical that the calibration model can accurately model visibilities from the baselines used in calibration. Relying on short baselines for calibration would require a highly accurate and complete model of diffuse foreground emission, since these short baselines are sensitive to large-scale structure on the sky. In the absence of a diffuse foreground emission model, it is advantageous to calibrate to \textit{long} baselines only \citep{Patil2016}.

\begin{figure}
\centering
\includegraphics[width=\columnwidth]{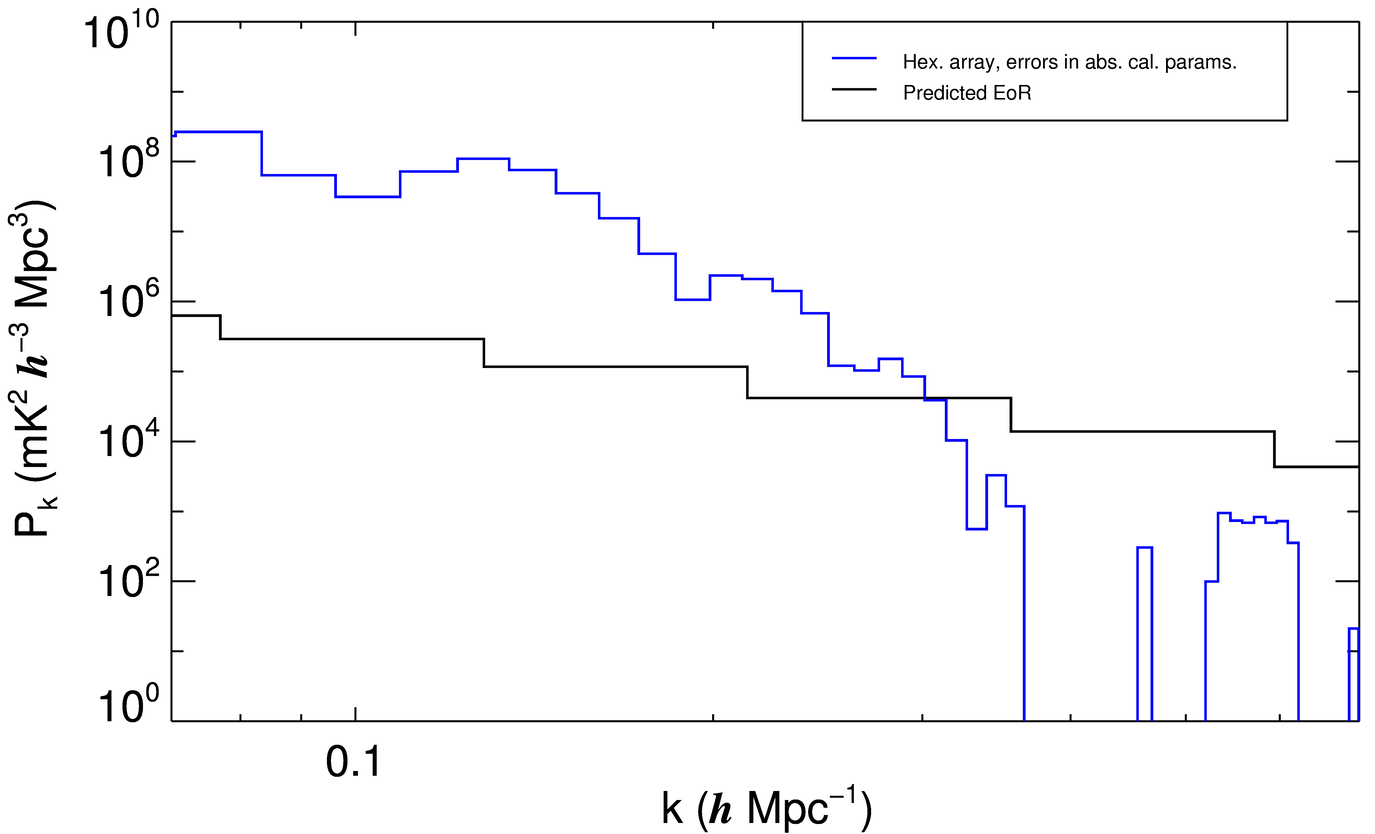}
\caption{1-D representation of power leakage in the EoR window due to errors in absolute calibration from sky model incompleteness. The simulations are based on the hexagonal array pictured in Figure \ref{fig: hex antenna layout large}. The blue line represents the difference in EoR window power between simulations with errors in absolute calibration and those with perfect calibration. The black rectangular outlines in Figure \ref{fig: 2d ps} indicate the 2-D PS modes that contribute to the blue line. The black line is the predicted EoR signal. Excess power in the EoR window due to absolute calibration errors overwhelm the predicted EoR signal on most of the PS modes plotted here.}
\label{fig: hex 1d ps}
\end{figure}

To quantify the power leakage in the EoR window due to absolute calibration errors from an incomplete sky model, we first subtract the perfect calibration PS in Figure \ref{fig: 2d ps}(d) from the PS with absolute calibration errors in Figure \ref{fig: 2d ps}(a). We then select a slice of the EoR window region of the 2-D PS that spans $k_{\parallel}=0.07-1.0 \ h \text{Mpc}^{-1}$ and $k_{\bot}=8.15\times10^{-3}-1.015\times10^{-2}\ h \text{Mpc}^{-1}$. This slice is centered on the mode measured by the shortest 15~m baselines in the array and is a characteristic representation of EoR window contamination that avoids regions of low UV coverage. The black rectangular outlines in the PS in Figure \ref{fig: 2d ps} delimit the region. Finally, we average over the power in this slice to produce a 1-D representation of power leakage in the EoR window as a function of PS mode $|k|$. We plot the result in blue in Figure \ref{fig: hex 1d ps}. The black line in Figure \ref{fig: hex 1d ps} is a fiducial EoR signal \citep{Furlanetto2006b}.

Figure \ref{fig: hex 1d ps} demonstrates that absolute calibration errors from sky model incompleteness can contaminate the PS measurement to an extent that dwarfs the EoR signal. The PS modes plotted here are characteristic of the EoR window, the region of the PS that is sensitive to the EoR. An EoR detection is not possible in modes in which foreground power leakage exceeds the EoR signal. The power leakage shown here is one error budget contribution corresponding to the limit of perfect array redundancy. Errors in relative calibration and thermal noise are not included in this simulation and will compound power leakage in the EoR window. An EoR detection will require mitigation of frequency-dependent absolute calibration errors to keep the total error budget within the required tolerances.


\section{Impact of Array Layouts on Calibration Errors} \label{section_layouts}

The magnitude of absolute calibration errors due to sky model incompleteness depends on array layout. Errors are driven by the frequency-dependent instrument PSF, which couples to sources that are missing from the calibration model. Arrays with good PSFs consolidate power in the true source locations, reducing the amount of power in the frequency-dependent source sidelobes.

Pseudo-random arrays sample the UV plane more uniformly than redundant arrays. In choosing an array layout, the benefits of redundant calibration must be weighed against the trade-offs associated with redundant arrays' degraded UV coverage. In this section we compare calibration errors for different classes of array layouts, some of which could support redundant calibration. We compare errors in the absolute calibration parameters due to calibration model incompleteness for each of these arrays.

\begin{figure*}
\centering
\includegraphics[width=2\columnwidth]{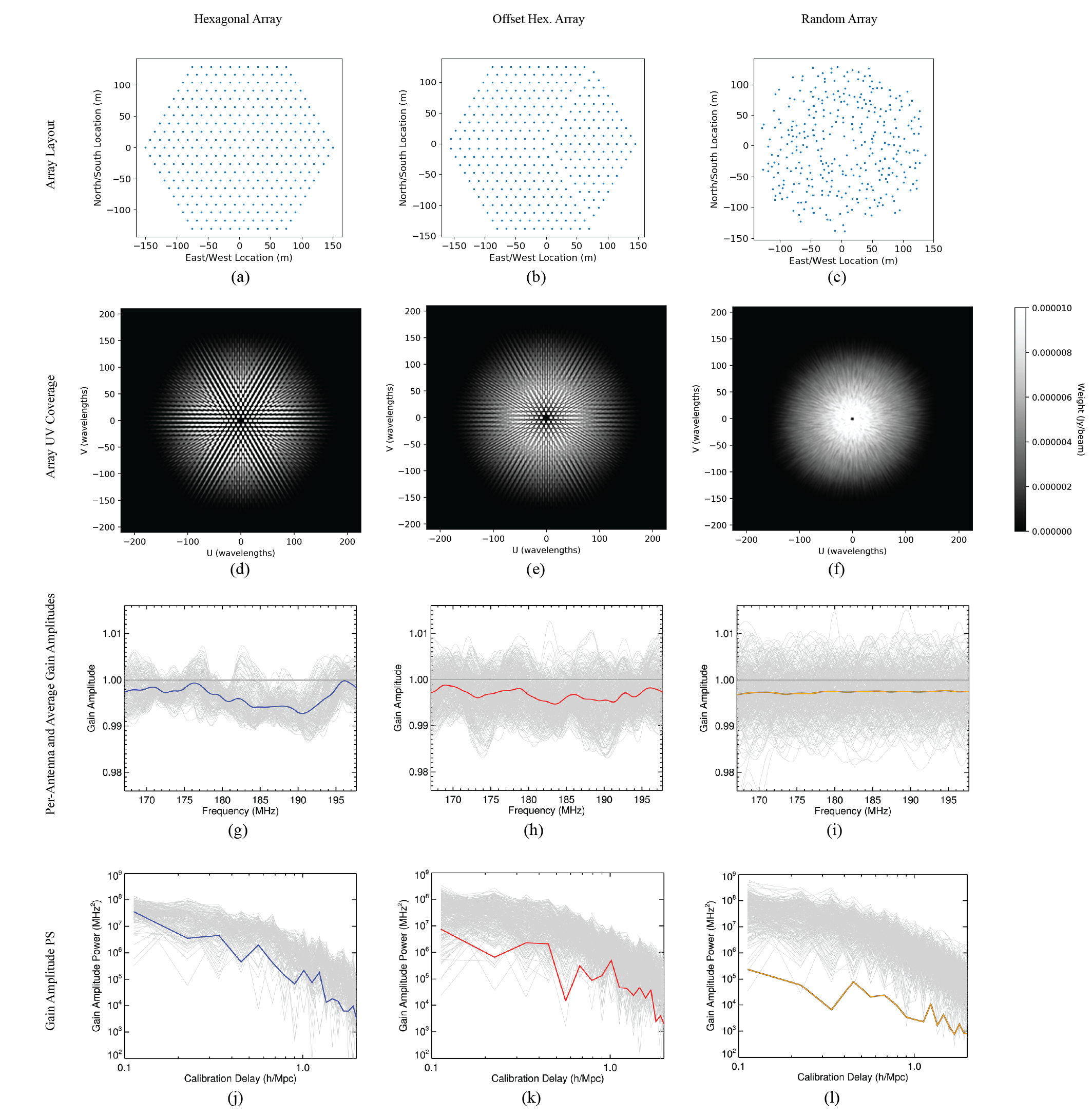}
\caption{Comparison of the gain amplitude errors for three array configurations. The top row shows the antenna positions for each configurations of 331 antennas: from left to right, a simple hexagonal array (also pictured in Figure \ref{fig: hex antenna layout large}), an offset hexagonal array, and a randomized array. The antenna locations in the randomized array are chosen to approximate the radial antenna distribution of the simple hexagonal array. The second row shows plots of the UV weights for each array configuration integrated across the 167-198 MHz frequency band and across a 2-minute observation. These plots are analogous to the measured response in the UV plane to a 1 Jy source at zenith. They illustrate the UV coverage and smoothness of each of the three arrays. The third row plots gain amplitudes as a function of frequency for each respective array. The grey lines represent the per-antenna gain amplitudes and the colored lines (also plotted in Figure \ref{fig: all arrays avg amp errors}) denote their average. The bottom row gives the PS (Fourier Transform squared) representation of the per-antenna and average gain amplitudes. Note that the random array has greater per-antenna gain amplitude variation than the other two array layouts but smaller average gain amplitude variation.}
\label{fig: per antenna array comparison}
\end{figure*}

To illustrate the impact of UV coverage on absolute calibration errors, we compare simulations of three array configurations, each with 331 antennas and with similar radial antenna distributions. The first is the simple hexagonal array discussed in \S\ref{section_errors} and pictured in Figures \ref{fig: hex antenna layout large} and \ref{fig: per antenna array comparison}(a). Next, we consider a hexagonal array divided into three sub-arrays offset by $\frac{1}{3}$ of the minimum antenna spacing (see Figure \ref{fig: per antenna array comparison}(b)) (\citealt{DeBoer2017}). This array configuration enables redundant calibration while offering better UV coverage than a simple hexagonal array. Finally, we simulate the randomized array pictured in Figure \ref{fig: per antenna array comparison}(c). To create this array, we calculate the radial baseline density of the hexagonal array from Figure \ref{fig: per antenna array comparison}(a) and randomly select 331 radial distances from that density distribution. We then randomly choose an azimuthal position for each antenna, requiring a minimum spacing of 5~m between antennas. For each the offset hexagonal array and the randomized array, we repeat the simulations described in \S\ref{section_errors}: we simulate visibilities from a 51,821-source catalog, calibrate to a 4,000-source sky model with the FHD software pipeline, and use those calibration solutions to calculate absolute calibration parameters from Equations \ref{A_calc} and \ref{grad_calc}.

\begin{figure}
\centering
\includegraphics[width=\columnwidth]{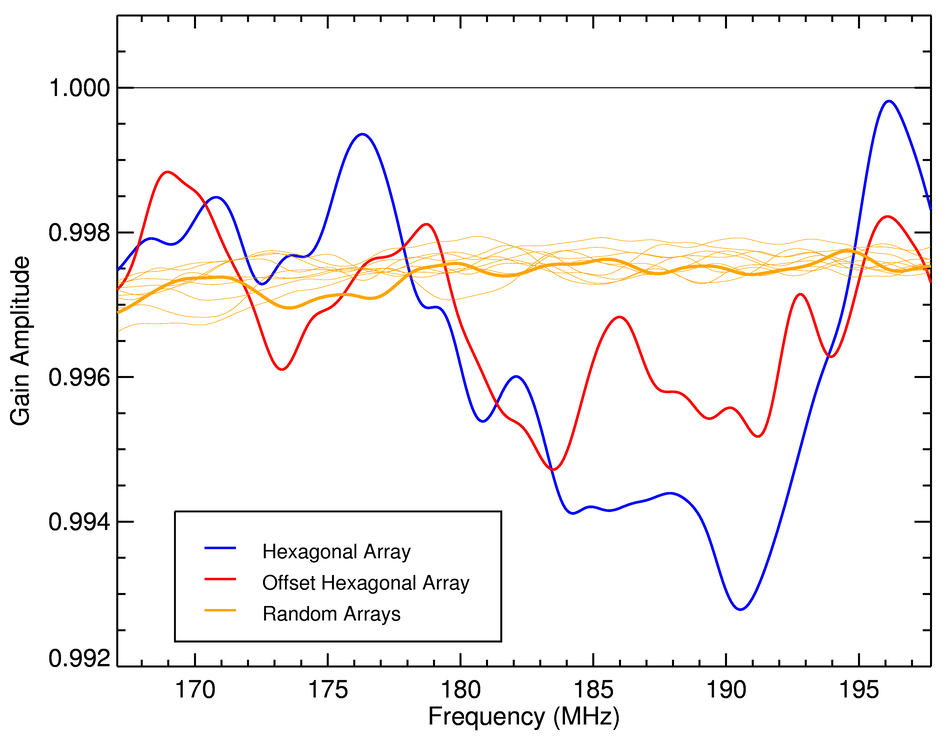}
\caption{Plot of the average gain amplitudes as a function of frequency, $
\hat{A}(f)$, for the hexagonal array in Figure \ref{fig: per antenna array comparison}(a) (blue), the offset hexagonal array pictured in Figure \ref{fig: per antenna array comparison}(d) (red), and ten realizations of random arrays (orange). The bold orange line corresponds to the random array pictured in Figure \ref{fig: per antenna array comparison}(c), and the additional faint orange lines illustrate the degree of variability across randomized realizations of the array configuration. The blue line is plotted alone in Figure \ref{fig: hex avg amp errors}. This figure illustrates that the random arrays have less variation in the average gain amplitude errors than the other two array configurations. The random arrays therefore exhibit less foreground power leakage in the EoR window region of the 2-D PS from absolute calibration errors.}
\label{fig: all arrays avg amp errors}
\end{figure}

Figure \ref{fig: all arrays avg amp errors} shows the average gain amplitudes $\hat{A}$ for each the simple hexagonal array (blue), the offset hexagonal array (red) and the randomized array pictured in Figure \ref{fig: per antenna array comparison}(c) (bold orange). The thin orange lines correspond to nine additional realizations of randomized arrays, illustrating the degree of variability expected across realizations. Figure \ref{fig: all arrays avg amp errors} indicates that randomized arrays have significantly smaller variations in $\hat{A}$ than either the hexagonal array or the offset hexagonal array, while the offset hexagonal array has slightly less $\hat{A}$ variation than the simple hexagonal array. Similarly, but not pictured here, the gain phase gradient parameters $\hat{\Delta}_x$ and $\hat{\Delta}_y$ have smaller variations for arrays with better UV coverage.

To understand this effect, it is helpful to consider the per-antenna sky-based calibration solutions alongside their average values. Figure \ref{fig: per antenna array comparison} plots the per-antenna gain amplitudes in gray for the simple hexagonal array (left column), the offset hexagonal array (middle column), and the randomized array (right column). The average gain amplitudes $\hat{A}$ are over-plotted in color. The third row plots the per-antenna and averaged gain amplitudes as a function of frequency. The bottom row gives the PS representation of the gain amplitudes by taking their Fourier Transform and squaring. This PS representation highlights the magnitude of the frequency structure in the gain amplitude errors and shows the power spectrum modes that are contaminated by these errors.

Although the randomized array's average gain amplitudes are less variable than those of either the simple or offset hexagonal arrays, its per-antenna gain amplitudes are actually more variable. This is because the error contributions for each antenna in a highly redundant array are correlated. An individual antenna's calibration solutions depend on that antenna's PSF, i.e. the PSF from all baselines that include the antenna. The hexagonal arrays have more uniform antenna PSFs than the random array and therefore have smaller per-antenna gain amplitude variations. However, modeled visibilities for redundant baselines experience exactly the same errors from missing sources in the calibration catalog. Thus errors in per-antenna calibration solutions will be correlated across any antennas that contribute to redundant baselines. When calculating the average calibration solutions across an array, errors average coherently in highly redundant arrays and incoherently in random arrays, leading to larger error variations in the absolute calibration parameters for the highly redundant arrays.

The impact of these errors in the absolute calibration parameters is apparent in the 2-D PS. Figure \ref{fig: 2d ps} presents 2-D PS for simulations of each the simple hexagonal array (left column), the offset hexagonal array (center column), and the random array realization pictured in Figure \ref{fig: per antenna array comparison}(c) (right column). As described in \S \ref{section_errors}, the top row of the figure shows the PS of simulated data calibrated with errors in the absolute calibration parameters. The bottom row shows the same data with perfect calibration. In each of these figures, the PS of the calibration model has been subtracted, leaving `residual' PS.

\begin{figure}
\centering
\includegraphics[width=\columnwidth]{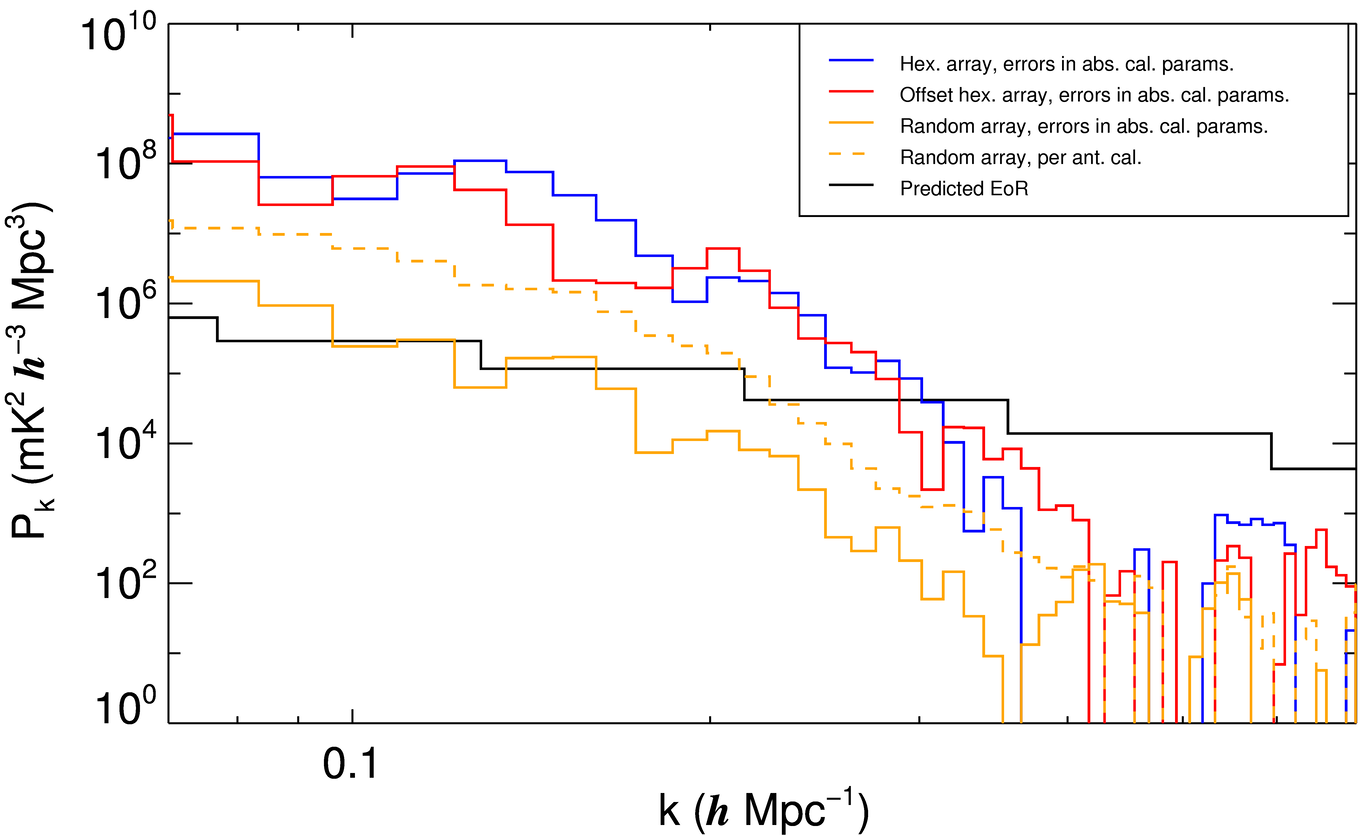}
\caption{1-D representation of power leakage from absolute calibration errors in the region of the EoR window indicated by the black rectangular outlines in Figure \ref{fig: 2d ps}. The solid blue, red, and orange lines are the difference in EoR window power between simulations with errors in absolute calibration and perfect calibration for each the simple hexagonal array (Figure \ref{fig: per antenna array comparison}(a)), the offset hexagonal array (Figure \ref{fig: per antenna array comparison}(b)), and the random array (Figure \ref{fig: per antenna array comparison}(c)). The blue line is also plotted in Figure \ref{fig: hex 1d ps}. The dashed yellow line represents the power leakage in the EoR window when, instead of allowing errors in only the absolute calibration parameters, we have implemented traditional sky-based calibration and allowed errors in every per-antenna, per-frequency calibration parameter. The black line is the predicted EoR signal.}
\label{fig: 1d ps}
\end{figure}

As in \S \ref{section_errors}, we produce 1-D plots of power leakage in the EoR window from the slices of the 2-D PS space delimited by the black rectangular outlines in Figure \ref{fig: 2d ps}. We subtract the perfect calibration PS from the PS that includes absolute calibration errors to isolate the power leakage that comes from the calibration errors. Next, we average the differences to represent power leakage as function of $|k|$. These results are plotted as the solid colored lines in Figure \ref{fig: 1d ps}. The blue line corresponds to the hexagonal array and is also plotted in Figure \ref{fig: hex 1d ps}. The red line corresponds to the offset hexagonal array, and the solid orange line corresponds to the random array from Figure \ref{fig: per antenna array comparison}(c). The black line is the fiducial EoR signal.

We also plot a dashed orange line corresponding to a simulation of the random array calibrated with a traditional per-antenna, per-frequency sky-based calibration scheme (i.e.\ minimizing Equation \ref{sky_cal_x2}). Under traditional sky-based calibration, sky model incompleteness introduces errors not only in the absolute calibration parameters but in all per-antenna, per-frequency calibration parameters. The discrepancy between the dashed and solid orange lines shows the magnitude of EoR window contamination from relative calibration errors for the random array.


\section{Discussion} \label{section_discussion}

Any calibration scheme that involves a sky model must contend with frequency-dependent calibration errors from sky model incompleteness. This includes redundant calibration techniques that eliminate sky model dependence in the relative calibration step but nonetheless require a sky model for absolute calibration. Error mitigation techniques must target and suppress frequency-dependent calibration errors. These techniques include developing near-perfect sky models, calibrating to short baselines, building instruments with extremely good UV coverage, and manufacturing antennas with very smooth bandpasses. The success of next-generation 21-cm cosmology experiments such as HERA and the SKA is contingent on their ability to sufficiently mitigate calibration errors.

Developing highly complete and accurate sky models is an active area of research \citep{Carroll2016, Hurley-Walker2017}. While the importance of sky model completeness has long been recognized in the context of sky-based calibration, redundant calibration also benefits from better sky modeling. Current efforts to image diffuse foreground structure will also enable better calibration of compact arrays that are sensitive to large-scale structure. While realistic sky models can never achieve perfect accuracy and completeness, better sky models minimize errors from the sky model incompleteness.

In \S\ref{section_errors} we explain that foreground power leakage from calibration errors falls off at a $k_{\parallel}$ threshold determined by the maximum baseline extent of the array. By calibrating to short baselines only, it may be possible to restrict contamination of the EoR window to low $k_{\parallel}$ modes \citep{Ewall-Wice2017}. These short-baseline calibration schemes are sensitive to the accuracy and completeness of the diffuse foreground model. To characterize these errors, future work must extend the simulations described in this paper to include diffuse emission.

In \S\ref{section_layouts} we show that errors from an incomplete sky model are reduced for arrays with more uniform UV coverage. For redundant arrays, improving UV coverage may mean using an array layout like the offset hexagon from Figure \ref{fig: per antenna array comparison}(d) \citep{DeBoer2017} instead of a simple hexagon like that from Figure \ref{fig: per antenna array comparison}(a) or adopting a hybrid array configuration with both redundant and non-redundant components such as the MWA Phase II \citep{mwa_phase_II_design}. Non-redundant arrays can have more uniform UV coverage than highly redundant arrays; the benefits of redundancy must be weighed against the greater errors in absolute calibration that result from poor UV coverage. Furthermore, arrays with sufficiently uniform antenna responses such that calibration benefits from averaging across antennas do not gain an advantage from redundant calibration. Averaging antennas eliminates the relative calibration degrees of freedom such that calibration consists of the absolute calibration step only. In that regime, array redundancy has an advantage if PS sensitivity, rather than calibration systematics, is a principal concern \citep{Parsons2012}.

Instruments with smooth spectral responses are the gold standard of precision calibration. Like antenna uniformity, spectral smoothness can be used as a prior on calibration solutions, prohibiting calibration errors from introducing power into the high $k_\parallel$ modes sensitive to the EoR \citep{trott_wayth_2016,deleraacedo_2017,trott_spectral_performance_2017}. However, if this prior is imposed on an instrument that is not inherently spectrally smooth calibration will not fit the true spectral features of the instrument response and they will contaminate the PS. To avoid contamination of the EoR signal, any spectral features in the antenna and receiver system faster than $\sim8$ MHz must be smaller than $\sim10^{-5}$ (\citealt{Barry2016}).


\section{Conclusion} \label{section_conclusion}

Redundant calibration requires a sky model to fit the absolute calibration parameters that are degenerate under relative calibration, as described in \S\ref{section_calibration}. Because of its sky model dependence, redundant calibration is susceptible to frequency-dependent calibration errors due to missing sources in the sky model, an effect identified by \citealt{Barry2016} in the context of sky-based calibration. Unless these errors are mitigated, this effect can quickly overwhelm the EoR signal, precluding a detection even in the limiting case of no noise and perfect redundancy.

In \S\ref{section_calibration} we present the mathematical framework of both sky and redundant calibration and decompose redundant calibration into two steps: relative calibration and absolute calibration. Using this framework, we extend the work of \citealt{Barry2016} to redundant calibration by exploring errors in absolute calibration in the limit of perfect relative calibration. This corresponds to a non-physical regime of perfect array redundancy and the absence of noise. In \S\ref{section_errors} we present results from simulations created with the FHD \citep{Sullivan2012} and $\epsilon$ppsilon \citep{Jacobs2016} software packages. We show that calibrating to an incomplete sky model introduces frequency-dependent errors and that exploiting array redundancy cannot eliminate these errors. Furthermore, we show that these errors can exceed the predicted EoR signal, precluding a detection. In \S\ref{section_layouts} we simulate these errors for several array configurations, showing that errors are suppressed for arrays with more uniform UV coverage.

A detection of the EoR will require mitigation of these calibration errors. Error mitigation can be accomplished by improving calibration models, using short-baseline calibration schemes, building arrays with good UV coverage, developing extremely spectrally smooth antenna and receiver systems, or through some combination of these strategies. Calibration errors are currently a dominant systematic limitation in the field and combating these errors should be a primary concern for the next generation of 21~cm cosmology arrays such as HERA and the SKA.

\section*{Acknowledgements}

This work was directly supported by NSF grants \#1613855, \#1506024, \#1835421, \#1613040, and \#1835421. Computation on the Amazon Web Services public cloud was supported by the University of Washington student-led Research Computing Club with funding provided by the University of Washington Student Technology Fee Committee. Parts of this research were supported by the Australian Research Council Centre of Excellence for All Sky Astrophysics in 3 Dimensions (ASTRO 3D), through project number CE170100013. A.P.B. is supported by an NSF Astronomy and Astrophysics Postdoctoral Fellowship under award \#1701440. C.M.T. is supported by an ARC Future Fellowship under grant FT180100196. The International Centre for Radio Astronomy Research (ICRAR) is a Joint Venture of Curtin University and The University of Western Australia, funded by the Western Australian State government. 

\bibliographystyle{apj}
\bibliography{main,library}
\end{document}